\documentclass[aps,pre,preprint,groupedaddress]{revtex4-2}
\usepackage{amsmath}
\usepackage{graphicx}
\usepackage{bm}
\usepackage{hyperref}
\usepackage{physics}
\usepackage{amsfonts}
\usepackage{amsmath}
\usepackage{amsthm}
\usepackage{lipsum} 
\usepackage{tikz}
\usetikzlibrary{decorations.pathreplacing}

\newtheorem{definition}{Definition}[section]
\newtheorem{remarkdef}{Remark}[definition]

\begin{document}

\title{Deriving friction force using fractional calculus}

\author{Georgii Koniukov}
\affiliation{University of Montpellier, 34090 Montpellier, France}

\begin{abstract}
The friction force is derived using fractional calculus by considering the non-uniform flow of time in dissipative processes. The approach incorporates inhomogeneous velocity without unphysical approximations, resulting in a Lagrangian where the order of fractional derivatives measures time intervals. The fractional term in the Lagrangian provides correct Euler-Lagrange, and ultimately, the Hamilton equations, and vanishes during energy change measurements, like a ghost.  
\end{abstract}

\maketitle

\tableofcontents
\newpage

\section{Introduction}
Since the mature formulation of the action principle by Euler, Hamilton, and Lagrange, it has been established that the equations of motion for dissipative linear dynamical systems with constant coefficients cannot be derived from a variational principle. In 1873, Lord Rayleigh introduced the Rayleigh dissipation function to handle the effects of velocity-proportional frictional forces in Lagrangian mechanics \cite{rayleigh1871}. In this situation, two scalar functions are required to define the equation of motion for a single system, which is not a very natural approach. Some limitations were rigorously demonstrated by Bauer in 1931 \cite{bauer1931}, who proved the impossibility of obtaining a dissipation term proportional to the first-order time derivative from such a principle. Over the years, various methods have been developed to address this issue, including the Bateman approach with auxiliary coordinates \cite{bateman1931} and time-dependent Lagrangians \cite{stevens1958}.
 Unfortunately, these methods often result in non-physical Lagrangians that fail to provide meaningful relations for the system's momentum and Hamiltonian. In 1975 Dekker \cite{dekker1975,dekker1981} introduces an innovative approach utilizing a Lagrangian that produces two first-order equations as complex conjugates, which is logical given the frequent appearance of complex numbers in the context of dissipation. These equations can be combined to yield a real, second-order equation of motion. However, it is important to note that this method is specifically applicable to a damped oscillator.

In 1996-1997, Riewe discovered a gap in Bauer's proof, which originally assumed that all derivatives were of integer order. Riewe incorporated Riemann-Liouville fractional time derivatives into the Lagrangian \cite{riewe1996}, successfully deriving equations of motion that included non-conservative forces such as friction. While innovative due to the redefinition of the action principle, this work relied on non-physical approximations and assumptions, such as inappropriately changing the limits of integration in fractional derivatives, applying an unphysical limit where the observation time interval is artificially set to zero, and selecting the fractional derivative order as $\frac{1}{2}$.

In 2014, Lazo and Krumreich refined Riewe's approach by replacing the Riemann-Liouville derivative with Caputo's fractional derivative and utilizing the semigroup property of fractional integral operators \cite{lazo2014}, which made the calculations more straightforward. However, the physical interpretation of fractional derivatives remained unclear. Additionally, issues such as the inappropriate modification of integration limits in fractional derivatives, the unphysical time limit, and the selection of the fractional derivative order as $\frac{1}{2}$ persisted.
\par Despite these advances, a unified and physically meaningful method to incorporate nonconservative forces within the framework of Lagrangian mechanics is still lacking. In this paper, we propose a novel approach that leverages the properties of fractional derivatives to derive nonconservative forces directly from a Lagrangian. Our method addresses the limitations of previous approaches by avoiding non-physical time limits and approximations, thereby providing a more natural and consistent description of dissipative systems. This new framework not only preserves the elegance of the variational principle but also enhances our understanding of the underlying physics in systems with friction and other nonconservative forces.

The structure of the paper is as follows: Section \ref{ii} introduces the necessary mathematical background on fractional calculus, focusing on the Riemann-Liouville and Caputo derivatives. Section \ref{iii} discusses the conceptual basis for modeling friction forces using fractional calculus. Section \ref{iv} introduces the Lagrangian for systems with fractional dissipative forces. Section \ref{v} presents the application of Hamilton's principle and derives the corresponding Euler-Lagrange equations. Section \ref{vi} derives the Hamilton equations for the system. Section \ref{vii} examines the changes in energy within the system. Finally, Section \ref{viii} discusses the physical implications of our findings and outlines potential directions for future research.

\section{The Riemann-Liouville and Caputo fractional calculus}\label{ii}
All formal mathematical definitions in this section have been adopted from the paper by Lazo and Krumreich \cite{lazo2014}.
\\
Let $t_0, t_1, t_1-t_0 \in \mathbb{R}_+$ and  $\rho(t) \in C^{[\alpha]+2}(\mathbb{R})$,  the notation $[.]$ denotes the ceiling function, which returns the smallest integer greater than or equal to the given value.

\begin{definition}[Integral operators]\label{fractionalIntegral}
\begin{equation}
_{t_0}I_t^\alpha \rho(t) = \frac{1}{\Gamma\left(\alpha\right)}\int\limits_{t_0}^t\frac{\rho(\tau)}{(t-\tau)^{1-\alpha}}d\tau \qquad \text{(left)} 
\end{equation}

and

\begin{equation}
_tI_{t_1}^\alpha \rho(t) = \frac{1}{\Gamma\left(\alpha\right)}\int\limits_t^{t_1}\frac{\rho(\tau)}{(\tau-t)^{1-\alpha}}d\tau \qquad \text{(right)} 
\end{equation}
\end{definition}

\begin{definition}[Riemann-Liouville fractional derivatives]
\begin{equation}\label{riemann}
_{t_0}D_t^\alpha \rho(t) = D_t^{[\alpha]+1}\  _{t_0}I_t^{1-\{\alpha\}} \rho(t)=\frac{d^{[\alpha]+1}}{dt^{[\alpha]+1}}\frac{1}{\Gamma\left(1-\{\alpha\}\right)}\int\limits_{t_0}^t\frac{\rho(\tau)}{(t-\tau)^{\{\alpha\}}}d\tau \quad \text{(left)} 
\end{equation}

and

\begin{equation}
_tD_{t_1}^\alpha \rho(t) = (-1)^{[\alpha]+1}D_t^{[\alpha]+1} \ _tI_{t_1}^{1-\{\alpha\}}\rho(t)= \frac{d^{[\alpha]+1}}{dt^{[\alpha]+1}}\frac{(-1)^{[\alpha]+1}}{\Gamma\left(1-\{\alpha\}\right)}\int\limits_t^{t_1}\frac{\rho(\tau)}{(\tau-t)^{\{\alpha\}}}d\tau \quad \text{(right)} 
\end{equation}
\end{definition}
\begin{definition}[Caputo fractional derivatives]
\begin{equation}
_{t_0}^CD_t^\alpha \rho(t) = \ _{t_0}I^{1-\{\alpha\}}_tD^{[\alpha]+1}_t\rho(t) = \frac{1}{\Gamma\left(1-\{\alpha\}\right)}\int\limits_{t_0}^t\frac{\rho^{([\alpha]+1)}(\tau)}{(t-\tau)^{\{\alpha\}}}d\tau \qquad \text{(left)} 
\end{equation}

and

\begin{equation}
_t^CD_{t_1}^\alpha \rho(t) = (-1)^{[\alpha]+1}\ _tI^{1-\{\alpha\}}_{t_1}D^{[\alpha]+1}_t\rho(t) = \frac{(-1)^{[\alpha]+1}}{\Gamma\left(1-\{\alpha\}\right)}\int\limits_t^{t_1}\frac{\rho^{([\alpha]+1)}(\tau)}{(\tau-t)^{\{\alpha\}}}d\tau \qquad \text{(right)} 
\end{equation}
\end{definition}
\begin{remarkdef}[Relations between Riemann-Liouville and Caputo fractional derivatives]\label{riemann-caputo}
\begin{equation}
_{t_0}^CD_t^\alpha \rho(t) = \ _{t_0}D_t^\alpha \rho(t) - \sum_{k=0}^{[\alpha]}\frac{\rho^{(k)}(t_0)}{\Gamma\left(k-\alpha+1\right)}(t-t_0)^{k-\alpha}
\end{equation}
and
\begin{equation}
_t^CD_{t_1}^\alpha \rho(t) = \ _tD_{t_1}^\alpha \rho(t) - \sum_{k=0}^{[\alpha]}\frac{\rho^{(k)}(t_1)}{\Gamma\left(k-\alpha+1\right)}(t_1-t)^{k-\alpha}
\end{equation}
\end{remarkdef}

\section{Conceptual basis for fractional friction forces}\label{iii}
According to the Noether's theorem, each symmetry in a physical system corresponds to a conservation law. In particular, the conservation of energy is a direct result of time invariance, meaning that the laws of physics are unchanged over time. This relationship is well-documented in classical mechanics and field theory, where the invariance of the Lagrangian with respect to time translations leads to energy conservation. Noether’s seminal paper, "Invariante Variationsprobleme" \cite{noether1918} laid the groundwork for these fundamental concepts, establishing the deep connection between symmetries and conservation laws in physics. However, when considering dissipative systems where energy is not conserved, time invariance must be broken. This can be achieved by introducing time-dependent elements or making the system explicitly time-inhomogeneous.  
\par Here by time inhomogeneity, we refer to introducing non-linearity in the time dependence of the system's parameters.  By breaking time invariance, these models should describe the transfer and dissipation of energy within a system, providing a more comprehensive understanding of real-world physical processes where ideal conservation laws do not strictly apply. 
\par Let's consider an evolution from time $t_0$ until $t_1$ in one-dimesional case. When time $\tau$ is linear one may try to find the position q(t) at any moment $t$ between $t_0$ and $t_1$ by integrating velocity $v$:
\begin{equation}\label{q(t_0)}
q(t)=\int\limits_{t_0}^tv(\tau)\mathrm{d}\tau+q(t_0)
\end{equation}
Or
\begin{equation}\label{q(t_1)}
    q(t)=q(t_1)-\int\limits_{t}^{t_1}v(\tau)\mathrm{d}\tau
\end{equation}
For dissipative systems one may consider non-linearity of time $(t-\tau)^\alpha$ (where $t$ shows dependence on the translation of time). For the first try we will consider only $0<\alpha<1$. One may postulate as well inhomogeneous velocity $\rho(\tau)$ to find the displacement $\Delta q(t)$ by integrating such velocity $\rho(\tau)$:
\begin{equation}
q(t)\propto\int\limits_{t_0}^t\rho(\tau)\mathrm{d}(t-\tau)^\alpha+q(t_0)
\end{equation}
However, since the Hamilton principle is evaluated in terms $\mathrm{d}\tau$, not $\mathrm{d}(t-\tau)^\alpha$:
\begin{equation}
q(t)\propto\int\limits_{t_0}^t\rho(\tau)\mathrm{d}(t-\tau)^\alpha+q(t_0)=-\alpha\int\limits_{t_0}^t\frac{\rho(\tau)}{(t-\tau)^{1-\alpha}}\mathrm{d}\tau+q(t_0)
\end{equation}
This reminds of Riemann-Liouville fractional integrals \ref{fractionalIntegral}. Then to have consistency with their mathematical properties we will choose the constant $\frac{1}{\Gamma(\alpha)}$ instead of $-\alpha$:
\begin{equation}\label{rho(t_0)}
q(t)-q(t_0)\equiv \ _{t_0}I_t^\alpha\rho(t)=\frac{1}{\Gamma(\alpha)}\int\limits_{t_0}^t\frac{\rho(\tau)}{(t-\tau)^{1-\alpha}}\mathrm{d}\tau
\end{equation}
Since $\rho(\tau)$ is inhomogeneous velocity, there could be an analogy with \ref{q(t_1)} for the $\tau>t$, and one might introduce as well ($0<\beta<1$ for the first try):
\begin{equation}\label{rho(t_1)}
    q(t_1)-q(t)\equiv \ _{t}I_{t_1}^\beta\rho(t)=\frac{1}{\Gamma(\beta)}\int\limits_t^{t_1}\frac{\rho(\tau)}{(\tau-t)^{1-\beta}}\mathrm{d}\tau
\end{equation}
And, thus,
\begin{equation}\label{timelink}
    q(t)=\ _{t_0}I_t^\alpha\rho(t)+q(t_0)=q(t_1)-\ _{t}I_{t_1}^\beta\rho(t)
\end{equation}
Let's remark on the consequences that should be true $\forall t \in [t_0;t_1]$ \ref{alpharemark}: 
\begin{equation}
    q(t_1)-q(t_0)= \ _{t_0}I_t^\alpha\rho(t)+ \ _{t}I_{t_1}^\beta\rho(t)
\end{equation}
The following sections are dedicated to the search for the values of $\alpha$ and $\beta$ \ref{alphabeta}.

\section{Lagrangian}\label{lagrangian_derivation}\label{iv}
Since $\rho(t)$ represents velocity, one may consider a Lagrangian where $\rho^2(t)$ serves as the dissipative component of the kinetic energy (representing energy loss) and $\frac{\gamma}{2}$ is an aspect ratio. We will adopt a quadratic form similar to Rayleigh \cite{rayleigh1871}. The negative sign preceding the "kinetic" term resembles the ghost term introduced by Lee and Wick \cite{lee1969}. However, in the work by Riewe \cite{riewe1996}, this term is complex. All these points emphasize that this term is likely non-measurable.

\begin{equation}
    L(t)=\frac{m\dot{q}^2(t)}{2}-\frac{\gamma \rho^2(t)}{2}-U(q)
\end{equation}
There is no energy loss at the start of observation, therefore $\rho(t_0)=0$. It is needed now to express $\rho(t)$ in terms of $q(t)$ to write the equations of motion later.
As we defined in \ref{rho(t_0)} (but with $q(t_0)=0$ as our frame of reference):
\begin{equation}
q(t)=\frac{1}{\Gamma(\alpha)}\int\limits_{t_0}^t\frac{\rho(\tau)}{(t-\tau)^{1-\alpha}}\mathrm{d}\tau
\end{equation}
Let's consider:
\begin{equation}
    ^C_{t_0}D^\alpha_t q (t)= \ ^C_{t_0}D^\alpha_t \ _{t_0}I_t^\alpha\rho(t)= \ _{t_0}I^{1-\{\alpha\}}_tD^{[\alpha]+1}_t \ _{t_0}I_t^\alpha\rho(t)
\end{equation}
Since $0<\alpha<1$:
\begin{equation}
     _{t_0}I^{1-\{\alpha\}}_tD^{[\alpha]+1}_t \ _{t_0}I_t^\alpha\rho(t)=\ _{t_0}I^{1-\alpha}_t\frac{d}{dt} \ _{t_0}I_t^\alpha\rho(t)
\end{equation}
According to \ref{riemann}:
\begin{equation}
     _{t_0}I^{1-\alpha}_t\frac{d}{dt} \ _{t_0}I_t^\alpha\rho(t)= \ _{t_0}I^{1-\alpha}_t \ _{t_0}D_t^{1-\alpha}\rho(t)
\end{equation}
According to \ref{riemann-caputo} and absence of dissipative kinetic energy at the beginning of observation ($\rho(t_0)=0$) we can change Riemann's fractional derivative of $\rho(t)$ with Caputo's:
\begin{equation}
     _{t_0}I^{1-\alpha}_t \ _{t_0}D_t^{1-\alpha}\rho(t)= \ _{t_0}I^{1-\alpha}_t \ _{t_0}^CD_t^{1-\alpha}\rho(t)= \ _{t_0}I^{1-\alpha}_t\ _{t_0}I^{1-\{1-\alpha\}}_tD^{[1-\alpha]+1}_t\rho(t)=\ _{t_0}I^{1-\alpha}_t\ _{t_0}I_t^\alpha\frac{d}{dt}\rho(t)
\end{equation}
According to the properties of Riemann-Liouville integrals \cite{samko1993} and the fact that $\rho(t_0)=0$:
\begin{equation}
    _{t_0}I^{1-\alpha}_t\ _{t_0}I_t^\alpha\frac{d}{dt}\rho(t)=\  _{t_0}I^1_t\frac{d}{dt}\rho(t)=\rho(t)
\end{equation}
Thus,
\begin{equation}\label{lagrangianCaputo}
    L(t)=\frac{m\dot{q}^2(t)}{2}-\frac{\gamma \left( \ ^C_{t_0}D^\alpha_t q (t)\right)^2}{2}-U(q)
\end{equation}
\section{Hamilton's principle and derivation of the friction force}\label{v}
In this chapter, we will derive the Euler-Lagrange equation for Hamilton's principle where Lagrangian \ref{lagrangianCaputo} depends on the Caputo Dearivatve \cite{agrawal2007}.
The action $S$ is:
\begin{equation}\label{action}
    S=\int\limits_{t_0}^{t_1}L\left(t,q(t),\dot{q}(t), \ ^C_{t_0}D^\alpha_t q (t)\right)\mathrm{d}t
\end{equation}
We will consider a case where we have initial condition $q(t_0)=0$.
\par Let $q(t)$ be the extremal function for the action \ref{action}. Then if we do the variation $q\mapsto q+\varepsilon\eta \ \left(\eta(t_0)=\eta(t_1)=0; \ \dot{\eta}(t_0)=\dot{\eta}(t_1)=0\right)$.
The condition for the extremum is obtained when the first Gâteaux variation is zero:
\begin{equation}
\delta S = \lim_{\epsilon \to 0} \frac{S[q + \epsilon \eta] - S[q]}{\epsilon} = \int_{t_0}^{t_1} \left[ \eta \frac{\partial L}{\partial q} + \dot{\eta} \frac{\partial L}{\partial  \dot{q}}  + \ ^C_{t_0} D_t^{\alpha}\eta\cdot\frac{\partial L}{\partial  \ ^C_{t_0} D_t^{\alpha}q}  \right] \mathrm{d}t = 0.
\end{equation}
Let's integrate by parts term by term ($\eta(t_0)=\eta(t_1)=0$):
\begin{equation}
    \int_{t_0}^{t_1}\dot{\eta} \frac{\partial L}{\partial  \dot{q}}\mathrm{d}t=\eta(t_1) \frac{\partial L}{\partial  \dot{q}}\Big|_{t=t_1}-\int_{t_0}^{t_1}\eta \frac{d}{dt}\frac{\partial L}{\partial  \dot{q}}\mathrm{d}t
\end{equation}
The term with fractional derivative:
\begin{equation}
    \int\limits_{t_0}^{t_1} \frac{\partial L}{\partial  \ ^C_{t_0} D_t^{\alpha}q}\cdot\ ^C_{t_0} D_t^{\alpha}\eta \  \mathrm{d}t= \int\limits_{t_0}^{t_1} \frac{\partial L}{\partial  \ ^C_{t_0} D_t^{\alpha}q}\cdot\ \left[\int\limits_{t_0}^t\frac{\dot{\eta}(\tau)\mathrm{d}\tau}{(t-\tau)^\alpha} \right] \mathrm{d}t
\end{equation}
We can change the order of integrals according to the integration domain \cite{love1937} and then integrate by parts:
\begin{figure}[h!]
    \centering
    \includegraphics[width=0.4\textwidth]{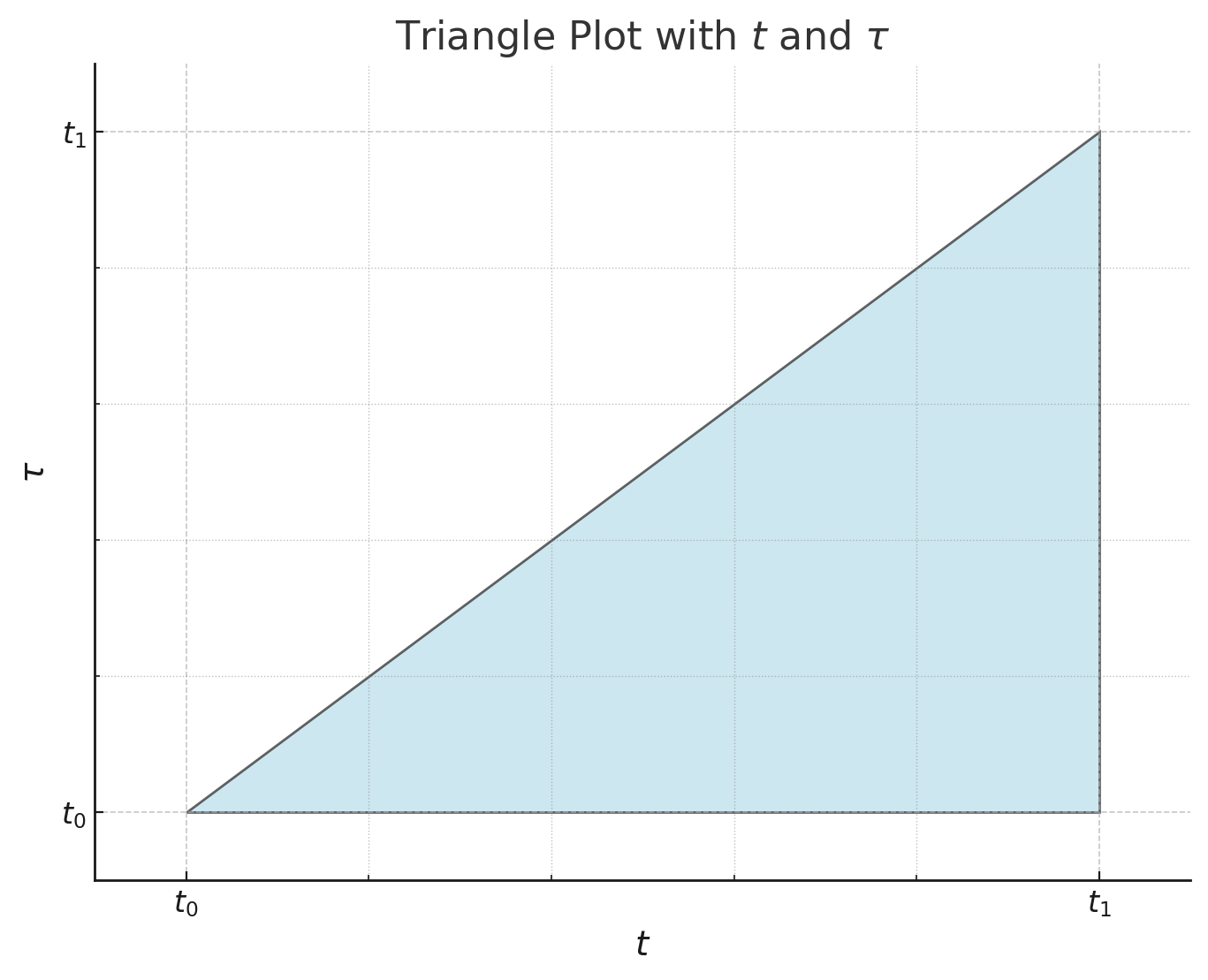}
    \caption{Integration domain}
    \label{fig:triangle_plot}
\end{figure}
\begin{equation}
   =\int\limits_{t_0}^{t_1} \dot{\eta}(\tau)\cdot\ \left[\int\limits_{\tau}^{t_1}\frac{\frac{\partial L}{\partial  \ ^C_{t_0} D_t^{\alpha}q}\mathrm{d}t}{(t-\tau)^\alpha} \right] \mathrm{d}\tau =\eta(\tau)\ \left[\int\limits_{\tau}^{t_1}\frac{\frac{\partial L}{\partial  \ ^C_{t_0} D_t^{\alpha}q}\mathrm{d}t}{(t-\tau)^\alpha} \right] \bigg|_{t_0}^{t_1}-\int\limits_{t_0}^{t_1} \eta(\tau)\cdot\frac{d}{d\tau} \left[\int\limits_{\tau}^{t_1}\frac{\frac{\partial L}{\partial  \ ^C_{t_0} D_t^{\alpha}q}\mathrm{d}t}{(t-\tau)^\alpha} \right] \mathrm{d}\tau
\end{equation}
Since $\eta(t_0)=0$, \ $\int\limits_{\tau}^{t_1}\frac{\frac{\partial L}{\partial  \ ^C_{t_0} D_t^{\alpha}q}\mathrm{d}t}{(t-\tau)^\alpha}\bigg|_{\tau=t_1}=0$ and \ref{riemann}:
\begin{equation}
    =\int\limits_{t_0}^{t_1} \eta(\tau)\cdot \ _\tau D^\alpha_{t_1}\frac{\partial L}{\partial  \ ^C_{t_0} D_t^{\alpha}q}\mathrm{d}\tau
\end{equation}
Thus,
\begin{equation}
    0=\delta S=\int_{t_0}^{t_1} \left[ \eta \frac{\partial L}{\partial q} -\eta \frac{d}{dt}\frac{\partial L}{\partial  \dot{q}} +\eta\cdot \ _t D^\alpha_{t_1}\frac{\partial L}{\partial  \ ^C_{t_0} D_t^{\alpha}q} \right] \mathrm{d}t
\end{equation}
And the Euler-Lagrange equation:
\begin{equation}\label{Euler-Lagrange}
    \frac{\partial L}{\partial q} - \frac{d}{dt}\frac{\partial L}{\partial  \dot{q}} +  \ _t D^\alpha_{t_1}\frac{\partial L}{\partial  \ ^C_{t_0} D_t^{\alpha}q}=0
\end{equation}
Let's consider the last term repeating the whole algorithm from \ref{lagrangian_derivation}:
\begin{equation}\label{friction_force}
   \gamma \ _t D^\alpha_{t_1} \ ^C_{t_0} D_t^{\alpha}q=\gamma \ _t D^\alpha_{t_1}\rho
\end{equation}
Let's calculate (remembering that Caputo derivative of a constant $q(t_1)$ is 0):
\begin{equation}
    ^C_tD^\beta_{t_1} q (t)= \ ^C_{t}D^\beta_{t_1}\left(q(t_1)- \ _{t}I_{t_1}^\beta\rho(t)\right)=- ^C_{t}D^\beta_{t_1}\ _{t}I_{t_1}^\beta\rho(t)
\end{equation}
Then by using \ref{riemann-caputo} (since $_{t_1}I_{t_1}^\beta\rho=0$):
\begin{equation}
    =-  _tD_{t_1}^\beta \ _{t}I_{t_1}^\beta\rho(t)=-\left(-\frac{d}{dt} \ _{t}I_{t_1}^{1-\beta} \ _{t}I_{t_1}^\beta\rho(t))\right)
\end{equation}
Again we use the semigroup property of Riemann-Liouville integrals \cite{samko1993}:
\begin{equation}
    =  \frac{d}{dt} \ _{t}I_{t_1}^{1}\rho(t)= \frac{d}{dt}\frac{1}{\Gamma\left(1\right)}\int\limits_t^{t_1}\frac{\rho(\tau)}{(\tau-t)^0}d\tau=\frac{d}{dt}\int\limits_t^{t_1}{\rho(\tau)}d\tau=-\rho(t)
\end{equation}
Thus, the equation \ref{friction_force} can be rewritten:
\begin{equation}
    \gamma \ _t D^\alpha_{t_1}\rho=-\gamma \ _tD^\alpha_{t_1}\ ^C _tD^\beta_{t_1} q (t)=-\gamma\left(-\frac{d}{dt} \ _{t}I_{t_1}^{1-\alpha}\left(-\ _{t}I_{t_1}^{1-\beta}\frac{d}{dt} \right)\right)q(t)=-\gamma\frac{d}{dt} \ _{t}I_{t_1}^{2-\alpha-\beta}\dot{q}(t)
\end{equation}
Now, it is possible to see the important restriction on $\alpha$ and $\beta$. Since the friction force should be proportional $\dot{q}(t)$ $\quad \Rightarrow \quad \alpha+\beta=1$:\label{alphabeta}
\begin{equation}
    =-\gamma\frac{d}{dt} \ _{t}I_{t_1}^{1}\dot{q}(t)=\gamma\dot{q}(t)
\end{equation}
And the Euler-Lagrange equation:
\begin{equation}
m\ddot{q}+\gamma\dot{q}+\frac{\partial U(q)}{\partial q}=0
\end{equation}
As a remark: adding \ $_{t_0}^CD_t^\alpha q \cdot \ _{t}^CD_{t_1}^{1-\alpha}$ term to the Lagrangian does not change the equation of motion like adding $\frac{df}{dt}$ where $f$ is arbitrary.
\section{Hamilton equations}\label{vi}
To obtain a physically meaningful Hamiltonian and corresponding equations of motion, it is essential to perform a Legendre transformation. In this context, we will treat the fractional derivative of $q$ as an independent projection, as it is not influenced by the velocity at the present time \( t \), but rather by the velocities at previous times. Remembering \ref{lagrangianCaputo}:
\begin{equation}
    L(q,q^{(\alpha)},\dot{q})=\frac{m\dot{q}^2}{2}-\frac{\gamma \left( q^{(\alpha)}\right)^2}{2}-U(q)
\end{equation}
In order to make H not dependent on $\dot{q},q^{(\alpha)}$ we will define canonical $p\equiv\frac{\partial L}{\partial \dot{q}},p_\alpha\equiv\frac{\partial L}{\partial q^{(\alpha)}}$:
\begin{equation}
    H(p,p_\alpha,q)\equiv p\dot{q}+p_\alpha q^{(\alpha)}-L(q,q^{(\alpha)},\dot{q})
\end{equation}
\begin{equation}
    dH=\dot{q}dp+q^{(\alpha)}dp_\alpha-\frac{\partial L}{\partial q}dq
\end{equation}
Meanwhile:
\begin{equation}
    dH=\frac{\partial H}{\partial p}dp+\frac{\partial H}{\partial p_\alpha}dp_\alpha+\frac{\partial H}{\partial q}dq
\end{equation}
Thus, according to \ref{Euler-Lagrange}:
\begin{equation}
    \begin{cases}
        \frac{\partial H}{\partial p}=\dot{q} \\
        \frac{\partial H}{\partial p_\alpha}=q^{(\alpha)}\\
        \frac{\partial H}{\partial q}=-\frac{d}{dt}p+ \ _tD_{t_1}^\alpha p_\alpha
    \end{cases}
\end{equation}
\begin{equation}
    H(p,p_\alpha,q)=\frac{p^2}{2m}-\frac{p_\alpha^2}{2\gamma}+U(q)
\end{equation}
And, actually, it is now evident that inhomogeneous velocity \ref{rho(t_0)} is a canonical momenta $p_\alpha$, and it, probably, should be $\alpha$-dependent \ref{alpharemark}:
\begin{equation}
    p_\alpha^2=(\gamma \rho)^2
\end{equation}
\section{Energy change}\label{vii}
With the inclusion of the new dissipative term in the Lagrangian, we will observe that, in this case, the energy and the Hamiltonian do not coincide.
\begin{multline}
    \frac{dH}{dt}=\frac{\partial H}{\partial p} \frac{dp}{dt}+\frac{\partial H}{\partial p_\alpha} \frac{dp_\alpha}{dt}+\frac{\partial H}{\partial q} \frac{dq}{dt}=\frac{\partial H}{\partial p} \frac{dp}{dt}+\frac{\partial H}{\partial p_\alpha} \frac{dp_\alpha}{dt}+\left(-\frac{dp}{dt}+ \ _tD_{t_1}^\alpha p_\alpha\right) \frac{dq}{dt}\\=\ ^C_{t_0}D^\alpha_t q \cdot\dot{p_\alpha}+\dot{q}\cdot\ _tD_{t_1}^\alpha p_\alpha 
\end{multline}
Here we can define the anticommutator:
\begin{equation}\label{dH/dt1}
    \frac{dH}{dt}= \{q,p_\alpha\}\equiv\ ^C_{t_0}D^\alpha_t q \cdot\dot{p_\alpha}+\dot{q}\cdot\ _tD_{t_1}^\alpha p_\alpha=-\frac{p_\alpha}{\gamma}\dot{p_\alpha}+\dot{q}(-\gamma\dot{q})
\end{equation}
On the other side we can introduce the energy $E$:
\begin{equation}
    H(p,p_\alpha,q)=\frac{p^2}{2m}-\frac{p_\alpha^2}{2\gamma}+U(q)\equiv E-\frac{p_\alpha^2}{2\gamma}
\end{equation}
\begin{equation}\label{dH/dt2}
     \frac{dH}{dt}=\frac{dE}{dt}-\frac{p_\alpha}{\gamma}\dot{p_\alpha}
\end{equation}
By comparing \ref{dH/dt1} and \ref{dH/dt2} we can obtain the energy dissipation law:
\begin{equation}
    \frac{dE}{dt}=-\gamma\dot{q}^2
\end{equation}

\section{Summary and Concluding Remarks on the Physical Significance}\label{viii}
\begin{itemize} 
   \item The concept of fractional derivatives lies in the non-uniform flow of time for dissipative processes, potentially offering a new understanding of the flow of time in physical processes. Philosophically, this can be understood as the different perceptions of time flow during energy expenditure/active engagement versus periods of inactivity.
    \item The use of Caputo fractional derivatives, both left and right, carries the same physical meaning; the difference lies in whether we determine the coordinate by knowing the future or the past. To define the Caputo fractional derivative, we integrate the velocity. The same applies to the usage of Riemann-Liouville derivatives, but we integrate the coordinate to define them.
    \item The dissipative term in the Lagrangian resembles "ghost term" \cite{lee1969} because it gives the correct Euler-Lagrange and Hamilton equations of motion, but disappears when we measure energy change.
    \item When calculating \( \frac{dH}{dt} \), natural commutation relations between canonical generalized coordinates arise, caused by integration by parts of the terms with fractional derivatives in the action principle.
    \item The condition that the sum of \( \alpha + \beta = 1 \), probably, means that the order of the fractional derivative is a measure of the time intervals of the past and future, respectively.  
    \begin{figure}[h!]
    \centering
\begin{tikzpicture}
    \draw[thick] (0,0) -- (10,0);
    
    \node at (0, -0.5) {$t_0$};
    \node at (10, -0.5) {$t_1$};
    \node at (3, -0.5) {$t$};
    
    \draw[dotted] (5,0) -- (5,-0.5);
    
    \node at (1.5, -0.75) {$\alpha$};
    \node at (6.5, -0.75) {$1-\alpha$};
    
    \draw[decorate,decoration={brace,amplitude=10pt,mirror}] (0,-0.1) -- (3,-0.1) node[midway,yshift=0.5cm] {};
    \draw[decorate,decoration={brace,amplitude=10pt,mirror}] (3,-0.1) -- (10,-0.1) node[midway,yshift=0.5cm] {};
\end{tikzpicture}
    \caption{Derivative's fractional order as a measure of time interval}
    \label{fig:time}
\end{figure}
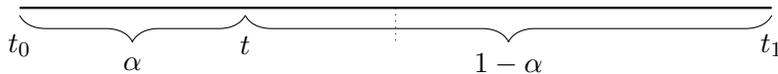
\item Since we did not consider $\alpha=\alpha(t)$ in the derivation, this, probably, means that for each time $t$ nature recalculates the Hamilton principle with new fixed $\alpha$. \label{alpharemark}
    \item It might be worth considering the Lagrangian as a function of the fractional derivative order $L(0,\alpha,1)$, with boundary values at zero and one, representing kinetic and potential energy, respectively.
    \item It is natural to continue the study of this method in quantum field cases and to rigorously define all concepts in the language of functional spaces and possibly interpret Planck's constant $i\hbar$ as a complex viscosity/damping coefficient, since it does not change the energy as the real viscosity coefficient $\gamma$, but stays with the first-order time derivative.
    \item  Furthermore, we hope that the insights gained from this study could enhance our understanding of the Navier-Stokes equations, which govern the motion of fluid substances. These equations, fundamental to fluid dynamics, describe how the velocity field of a fluid evolves over time under the influence of various forces, including viscosity \cite{stokes1880}. 
    \end{itemize}

\end{document}